\nonstopmode

\documentclass[prl,twocolumn,showpacs,preprintnumbers]{revtex4}
\usepackage{graphicx}
\usepackage{dcolumn}
\usepackage{amsmath}

\newcommand{\beq}{\begin{equation}}
\newcommand{\eeq}{\end{equation}}
\newcommand{\beqa}{\begin{eqnarray}}
\newcommand{\eeqa}{\end{eqnarray}}

\def\Im{\hbox{Im}}
\def\Re{\hbox{Re}}

\def\latt{\text{latt}}

\begin{document}

\title{ Dynamical Breakup of the Fermi Surface in a doped Mott Insulator}
\author{M. Civelli$^1$, M. Capone $^2$ $^3$, S. S. Kancharla $^4$, O. Parcollet $^5$ and  G. Kotliar$^1$}
\affiliation{$^1$ Physics Department and Center for Materials
Theory, Rutgers University, Piscataway NJ USA}
\affiliation{ $^2$ INFM-SMC and Istituto dei Sistemi Complessi CNR,
Via dei Taurini 19, I-00185, Rome, Italy}
\affiliation{$^3$ Physics Department, University of Rome ``La
Sapienza'',  Piazzale A. Moro 5, I-00185, Rome, Italy}
\affiliation {$^4$ Departement de physique and
Regroupement quebecois sur les materiaux de pointe, Universit\'e
de Sherbrooke, Sherbrooke, Quebec J1K 2R1, Canada}
\affiliation{$^5$
Service de Physique Theorique, CEA Saclay , 91191, Gif-Sur
Yvette, France}

\begin{abstract}
The evolution from an anomalous metallic phase to a Mott insulator
within the two-dimensional Hubbard model is investigated by means of
the Cellular Dynamical Mean-Field Theory.
We show that the density-driven Mott metal-insulator
transition is approached in a non-uniform way
in different regions of the momentum space. This gives rise to
a breakup of the Fermi surface and to the formation of {\it hot}
and {\it cold} regions, whose position depends on the hole or electron like
nature of the carriers in the system.
\end{abstract}

\date{\today}
\maketitle

The mechanism of high-temperature superconductivity
remains mysterious after nearly two decades of
research. Among many theoretical proposals,
a line of thought stresses  the  importance of the proximity to a Mott
insulating state and of understanding the  anomalous metallic state
originating from doping
this Mott insulator, a state which is not described by Fermi-liquid
theory \cite{anderson}.
Along this path two very different
approaches have been implemented: variational wavefunction studies
\cite{gross-arun} and slave boson approaches \cite{bza-liu}.
These methods allow to connect the Mott insulator to an anomalous normal
state, interpreting the phenomenon of high temperature superconductivity
as a direct consequence of the proximity
to the Mott Metal-Insulator Transition (MIT).
Their early successes include the prediction of the
d-wave symmetry of the order parameter, the existence of a pseudogap
close to half-filling and the dome-like shape of the critical
temperature.

In this letter we investigate this issue using the recently developed
Cellular Dynamical Mean-Field Theory (CDMFT)\cite{cdmft},
applied to the two-dimensional Hubbard model on the square lattice.
The CDMFT method reduces a quantum many body problem to a cluster of sites
embedded in an
effective medium described by a Weiss function.
This quantity is determined self-consistently and can
accommodate various broken symmetries, such as antiferromagnetism
or $d$-wave superconductivity. In this work we consider
a metallic phase which does not break any symmetry
and follow its evolution as
a function of parameters, such as the doping with respect to the
Mott insulator. States with long-range order, including superconductivity and
antiferromagnetism \cite{jarrellsc-lich}, will be discussed in a following paper. CDMFT offers the advantage of
treating  both the coherent and
the incoherent excitations  (quasiparticle peak and Hubbard bands) on the same
footing, capturing the short-range physics of singlet formation on
links, using self-energies which provide a dynamical
(i.e. frequency dependent) generalization of the condensates of
the early slave boson mean field theory.
The Hubbard model Hamiltonian is:
\begin{eqnarray}
H = -\sum_{ i,j,\sigma} t_{ij}\, (c^{\dagger}_{i,\sigma} c_{j,\sigma} +
h.c.) + U \sum_i n_{i\uparrow}n_{i\downarrow} -\mu\sum_i n_i \nonumber
\label{hamiltonian}
\end{eqnarray}
where $c_{i,\sigma}$ ($c^{\dagger}_{i,\sigma}$) are destruction (creation)
operators for electrons of spin $\sigma$, $n_{i\sigma}$ is the density of
$\sigma$-spin electrons, $U$ is the on-site
repulsion and $\mu$ the chemical potential which determines the
electron density $n=1/N_c \sum_{i\sigma} \langle c^{\dagger}_{i\sigma}
c_{i\sigma} \rangle$, $N_c$ is the total number of sites.
The hopping amplitude
$t_{ij}$ is chosen to be
limited to nearest-neighbors $t$ and to next-nearest-neighbors $t^{\prime}$,
and a strong repulsion $U= 16t$ is assumed.
In the following we will study the hole-doped system ($n<1$)
with $t^{\prime}=-0.3 t$, and the electron-doped system with
$t^{\prime}=+0.3 t$ and $n<1$ (equivalent to the $n>1$ system upon a particle-hole transformation which
reverses the sign of $t^{\prime}$). Our main results can be summarized as follows:

{\it i)} the normal state approaching the MIT exhibits the
phenomenon of momentum space differentiation \cite{damasce-campu}, similar to the results
of \cite{bpk}.
Different regions of momentum space behave very
differently : {\it hot} (resp. {\it cold}) regions are characterized by a
small (resp. large) quasiparticle residue $Z_k$ and a large (resp. small) inverse
scattering rate $\Im \Sigma $.
This effect gets most pronounced close to the MIT point.

{\it ii)} while the features described in {\it i)} are characteristic of proximity to
the MIT, their detailed mechanism and the location in $k$-space of the cold and hot
regions depend on the band parameter $t^{\prime}/t$.
In particular we find that the sign of $t^{\prime}/t$ for hole-like
\lbrack electron-like\rbrack \, doped cuprates corresponds to a location of the
cold region in the $k$-space around $({\pi \over 2},{ \pi \over 2}) $
\lbrack $(0, \pi ),\, (\pi, 0)$\rbrack. This finding agrees with experimental
observations \cite{damasce-campu} and earlier theoretical studies
\cite{jarrell,tohyama4,kusko,tremblay,tremblay2}.
Our study is complementary, but consistent, with direct exact
diagonalization studies or cluster perturbation theory on large clusters \cite{tremblay}.
Introducing the CDMFT self-consistency condition allows for a continuous set of
dopings and the high resolution in frequency needed to study the formation
and destruction of quasiparticles, as demonstrated in single-site
Dynamical Mean-Field Theory (DMFT) \cite{revmodmft} and, with remarkable success,
in tests in one dimension \cite{marce}.
In this paper, we argue that the proximity of the
MIT is central in this phenomenon and we investigate
this connection in detail.

\begin{figure}[!htb]
\includegraphics[width=6.0cm,height=2.0cm,angle=-0] {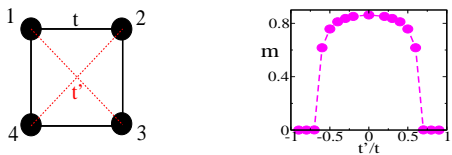}
\caption{Left side: the CDMFT Cluster.
Right side: staggered magnetization
as a function of the next-nearest hopping $t^{\prime}$ at half-filling.
The parameter $t^{\prime}$ controls the magnetic frustration in the system.}
\label{Fcluster}
\end{figure}

\begin{figure}[]
\includegraphics[width=8.0cm,height=8.75cm,angle=-0] {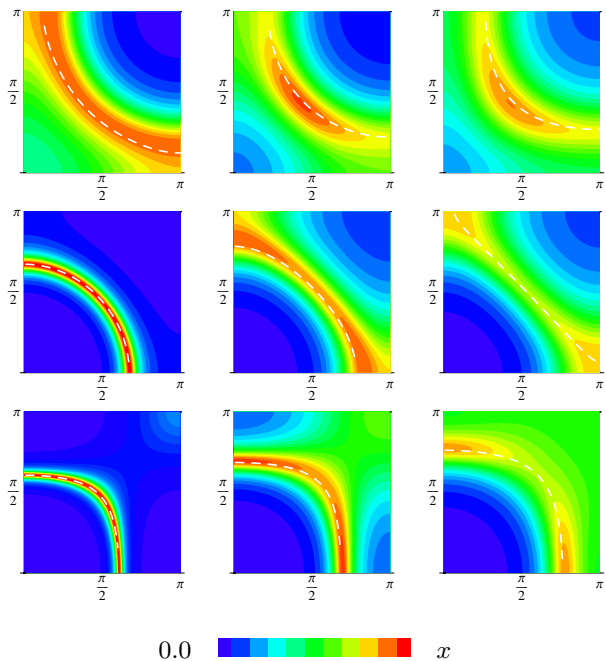}
\caption{$A(k,\omega=0^{+})$ in the first quadrant of the Brillouin zone.
From the top: in the first row
$t^{\prime}= -0.3 t$, $n = 0.73, 0.89, 0.96$, color scale $x= 0.28, 0.22, 0.12$;
in the second $t^{\prime}= +0.3 t$, $n = 0.70, 0.90, 0.95$,
color scale $x= 0.82, 0.34, 0.27$;
in the lowest row  $t^{\prime}= +0.9 t$, $n = 0.69, 0.92, 0.96$,
color scale $x= 0.90, 0.32, 0.22$. The white dashed line is the FS given
by $t_{\text{eff}}(k)=\mu$.
 }
\label{fA}
\end{figure}

As mentioned above, in CDMFT we have to solve for a cluster of sites
embedded in a self-consistent non-interacting bath.
Our cluster is a  $2\times 2$ plaquette (Fig. \ref{Fcluster}),
and the bath is truncated to 8 sites, in order to solve the
impurity model by means of a variant of the Exact
Diagonalization (ED) algorithm used in single-site DMFT \cite{krauth-caffarel}.
For the Matsubara frequency calculations we use
an effective inverse temperature \cite{krauth-caffarel}, which we set
equal to $\beta=128$ in units of the half bandwidth $4 t$.
The computation produces the self-energies $\Sigma_{ij}$ on the plaquette and
the lattice self-energy given by (using the square symmetry):
\begin{multline}
\label{self}
\Sigma_\latt(k,\omega)=  \Sigma_{11}(\omega)+ \Sigma_{12}(\omega) (\cos k_{x}+ \cos k_{y})+
\\
                    \Sigma_{13}(\omega)\, \cos k_{x} \cos k_{y}
\end{multline}
\begin{figure}[!htb]
\includegraphics[width=6.5cm,height=13.0cm,angle=-0] {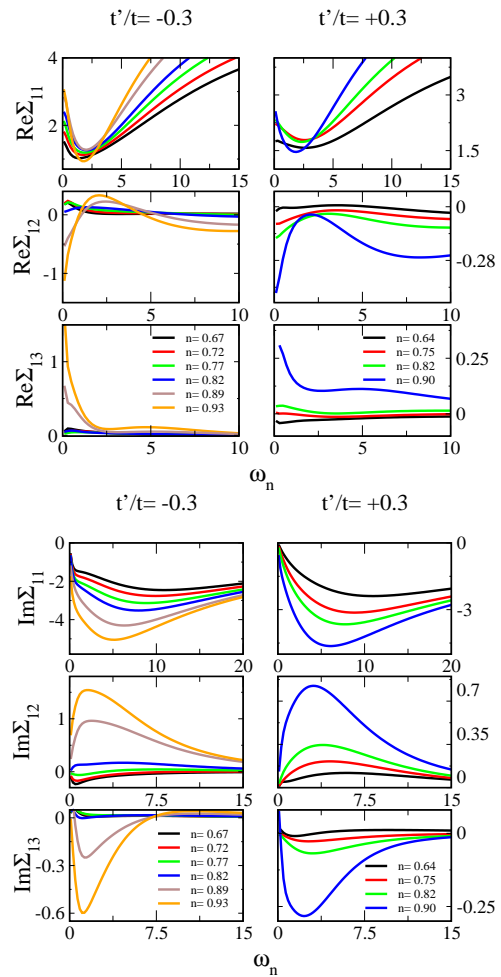}
\caption{Real and Imaginary part of the cluster self-energies for
 $t^{\prime}/t= \pm 0.3$. Notice the strong enhancement of the
nearest-neighbor an next-nearest-neighbor cluster self-energies
$\Sigma_{12}$ and $\Sigma_{13}$ close
to the MIT}
\label{fImS}
\end{figure}
The cluster self-energies on the imaginary frequency axis
(where the Green's functions of our ED discrete system are
smooth \cite{marce}) are presented in Fig. \ref{fImS}.
At large doping, we recover single-site DMFT results:
$\Sigma_{12},\Sigma_{13}\simeq 0$ and only
$\Sigma_{11}$ appreciably different from zero. The
lattice self-energy $\Sigma_\latt(k,\omega)$ (\ref{self})
is thereby $k$-independent.
However the cluster self-energies increase sizably at low doping as we
get close to the MIT and $\Sigma_\latt$ becomes
$k$-dependent.
\begin{figure}[!htb]
\begin{center}
\includegraphics[width=5cm,height=6cm,angle=-0] {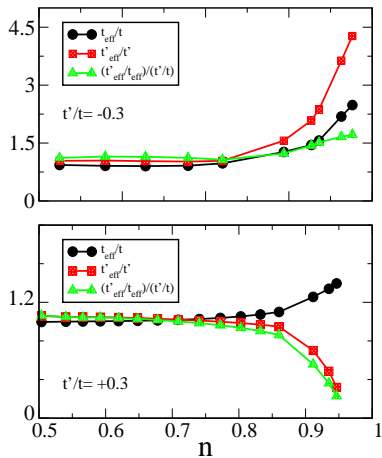} \quad
\end{center}
\caption{Renormalization of the  hopping coefficients
and of their ratio
as a function of density for $t^{\prime}/t=\pm 0.3$.}
\label{ft10.9}
\end{figure}
The zero frequency limit of the real part of the self-energy
determines the shape of the interacting Fermi Surface (FS), which
is given by $t_{\text{eff}}(k)=\mu$, where
$t_{\text{eff}}(k)\equiv \, t(k)-\Re \Sigma_{\text{latt}}(k,\omega=0^{+})$,
being $t(k)$ the Fourier transform of the hopping amplitude $t_{ij}$.
We find that the renormalization of the FS
becomes appreciable close to the MIT.
The self-energy itself depends weakly on the sign of $t^{\prime}$,
and in particular it has the same sign for both positive and negative
$t^{\prime}$.
However, given its large magnitude, when combined with
$t^{\prime}$ of different signs, it produces interacting FS's of
very different shape in the electron-doped and hole-doped case
(see Fig. \ref{fA}). This can be understood in terms  of  the renormalized low
energy hopping
coefficients $t_{\text{eff}}= t- \Re\Sigma_{12}(\omega=0^{+})/2$ and
$t^{\prime}_{\text{eff}}= t^{\prime}- \Re\Sigma_{13}(\omega=0^{+})/4$
presented in Fig. \ref{ft10.9} (where the $\omega=0^{+}$ limit is extrapolated
from the lowest Matsubara frequencies).
Correlations act to {\it increase } the value of
$t_{\text{eff}}$.
$\Re\Sigma_{13}$ is  negative, and it affects $t^{\prime}_{\text{eff}}$ quite
differently for $t^{\prime}/t=\pm 0.3$ . In fact it
increases  $t^{\prime}_{\text{eff}}$ in absolute value for $t^{\prime}=-0.3 t$,
in such a way that also the ratio $(t^{\prime}_{\text{eff}}/t_{\text{eff}})/
(t^{\prime}/t)$ weakly increases approaching the MIT,
thereby enhancing the hole-like curvature of the FS.
This has the effect of flattening horizontally the Fermi line in the
region of the momentum space close to $(0,\pi)$ or $(\pi,0)$. Because
of the enhancement in $t_{\text{eff}}$ in the vicinity
of the Mott transition point, in this region the Fermi line is
also weakly doping dependent. This behavior of the antinodal region
of cuprates superconductors in the normal state has been observed
experimentally \cite{Shen}.
On the other hand, $\Re \Sigma_{13}$
decreases  $t^{\prime}_{\text{eff}}$ for  $t^{\prime}=+0.3 t$, driving the FS
towards a flatter, nested FS as half-filling is approached.
This is clearly seen in Fig. \ref{fA} where the spectral function at zero
frequency  is presented in the first quadrant of the Brillouin Zone.

\begin{figure}[!htb]
\begin{center}
\includegraphics[width=8cm,height=8cm,angle=-0] {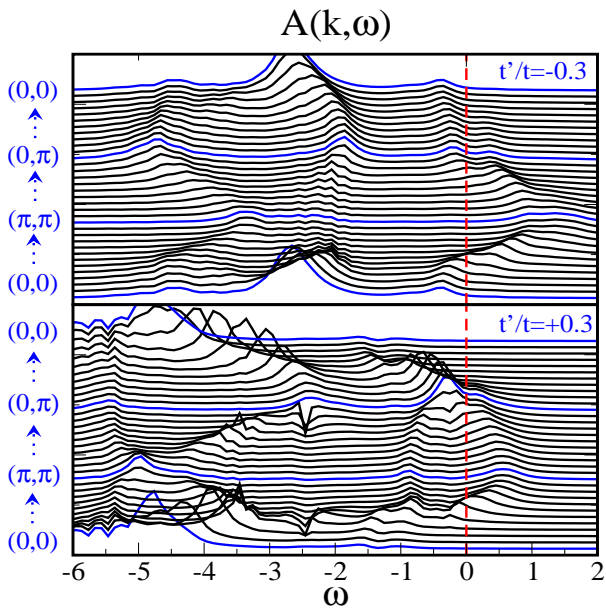}
\end{center}
\caption{Spectral function $A(k,\omega)$ as a function of $\omega$
along the path $(0,0)\rightarrow (\pi,\pi)\rightarrow (0,\pi)
\rightarrow (\pi,\pi)$. In the upper panel $t^{\prime}/t=-0.3$ and $n=0.96$,
in the lower panel $t^{\prime}/t=+0.3$ and $n=0.95$. The upper Hubbard band is
out of the energy scale covered by the figure.}
\label{FAw}
\end{figure}

Let us now turn to the formation of hot and cold regions.
For both cases $t^{\prime}/t=\pm 0.3$,  $\Im\Sigma_{11}$ and $\Im\Sigma_{13}$
have the same  negative
sign in contrast with the half-filled case of earlier studies on an
anisotropic model \cite{bpk}. While at $T=0$, $\Im \Sigma_{ij} (\omega=0) =0 $,
the finite effective temperature used in the ED calculation
gives rise to lifetimes which are strongly dependent on the
position on the FS and reflect the variation of the
quasi-particle residue $Z_k = \left( 1 -
\left. \frac{\partial\Im\Sigma_{\text{latt}}(k,i\omega_n)}{\partial\omega_{n}}\right|_{\omega_{n}\rightarrow 0}
\right)^{-1}$.
The slopes of all the imaginary parts of cluster self-energies increase as we
approach the Mott transition (see Fig. \ref{fImS}).
As they sum up differently to form $Z^{-1}_k$
in different regions of the $k$-space, according to the $k$-dependent
coefficient multiplying them, $Z_k$ may assume different values
on different points of the FS. This can be already seen in
Fig. \ref{fA} and \ref{FAw}.

In the case ${\bf t^{\prime}=+0.3 t}$ (electron-doped), as remarked above,
the FS is flat and crosses ($\pi/2,\, \pi/2$).
There $\Im \Sigma_\latt(k) \simeq \Im \Sigma_{11}$.
On the other hand, near  $(\pi, 0)$ \lbrack\,$(0, \pi )\,$\rbrack,
$\Im \Sigma_\latt(k)\simeq \Im \Sigma_{11} -\Im \Sigma_{13}$ .
Since $ \Im \Sigma_{11}$ and $\Im \Sigma_{13}$ have the same
sign,  we see that, while the local self-energy $\Sigma_{11}$
tends to decrease the quasiparticle residue and increase the
inverse scattering rate while approaching the Mott transition, its
effect is counterbalanced by  the growth
of $\Im \Sigma_{13}$, resulting in the formation of a cold patch
around $(0 ,\pi)$ \lbrack \,$(\pi, 0)$\,\rbrack.
For ${\bf t^{\prime}=-0.3 t}$ instead the effect of the real
part of  the self-energy is to bend the Fermi
line slightly away from $(\pi/2,\, \pi/2)$. The intersection of
the Fermi line  and the zone diagonal now occurs at  $(0.4 \,
\pi,\, 0.4\, \pi)$. At that point  $\cos k_{x}+ \cos k_{y}\simeq
\, 0.6$, while $\cos k_{x}\, \cos k_{y}$ is numerically  much
smaller ($\sim 0.1$). Hence $\Im \Sigma_\latt \simeq \Im
\Sigma_{11}+\, 0.6\, \Im \Sigma_{12} + 0.1\, \Im \Sigma_{13}$.
The growth of  $\Im \Sigma_{12}$ (which has the  opposite  sign
to $\Im \Sigma_{11}$ and $\Im \Sigma_{13}$) {\it reduces } the
strong correlation effects along the diagonal, and produces a
cold patch. This "screening" of the correlations is larger
than the one that takes place  near the
$(0, \pi)$ \lbrack \,$(\pi, 0)$\,\rbrack \, region,
where the screening is given by $\Im \Sigma_{13}$.
The modulation in $k$-space becomes in fact noticeable as soon as $ \Im
\Sigma_{12} \ge 1.7 \, \vert \Im \Sigma_{13} \vert $.
We argue that this effect is a precursor  to the
localization of particles that takes place at  MIT. The hot
quasiparticles localize, showing an insulator-like nature, while the cold
quasiparticles retain their dispersion as the Mott insulator is approached.
This can be
seen also in Fig. \ref{FAw}, where we  show the spectral weight $A(k,\omega)$
(at $4\%$ doping for $t^{\prime}/t= -0.3$, $5\%$ for $t^{\prime}/t=+0.3$)
as a function of the energy for different points in the Brillouin
zone, along the path $(0,0)\rightarrow (\pi,\pi)\rightarrow
(0,\pi) \rightarrow (0,0)$. In the case $t^{\prime}/t= -0.3$ (the upper
panel), a quasiparticle peak  disperses along the
$(0,0)\rightarrow (\pi,\pi)$, where a cold spot is formed at the
Fermi level; on the contrary the peak is less dispersive passing through the region
around $(0,\pi)$, where there is the hot spot. On the other hand, in the case $t^{\prime}/t=+0.3$
(lower panel) the quasiparticle peak is dispersing more in the vicinity
of the region $(0,\pi)$ and, while dispersing less, evaporates around $(\pi/2,\pi/2)$.
The hot-cold spots are in this way switched.

For $t^{\prime}=+0.3 t$ the Mott insulating state has antiferromagnetic
long range order.
To investigate the influence of antiferromagnetic correlation in
determining the evolution of the electronic structure
described above,
we have studied  (see  Fig. \ref{ft10.9}) the model with
a large $t^{\prime}=+0.9 t$. For this value of $t^{\prime}$,
in fact, we
can rule out any antiferromagnetic long-range order already at
half-filling. The right side of Fig. \ref{Fcluster} displays
the staggered magnetization of the system at half-filling,
$m= \sum_i (-1)^{i} (n_{i\uparrow}-n_{i\downarrow})/N_c$,
as a function $t^{\prime}/t$,
from $t^{\prime}/t= -0.9$ to $t^{\prime}/t=+0.9$: a sudden drop
of the magnetization is evident around $|t^{\prime}/t|\simeq 0.75$. The parent insulator
state in this case is likely to have valence-bond or flux
order \cite{affgk1}.
Despite the different nature of the parent insulator, the results for
$t^{\prime}=+0.9 t$ are similar to those for  $t^{\prime}=+0.3 t$.
This, however, does not exclude that qualitative differences may appear
between big and small $t^{\prime}$
with a better resolution in $k$-space (i.e. a bigger cluster).

To summarize, using Cellular DMFT  we have studied the MIT in
the Hubbard model.
The approach to the insulator is not Fermi-liquid-like, and anisotropic
properties appear in the momentum space. In particular, close to the MIT
there is a breakdown of the FS and the appearance of
{\it cold} and {\it hot} regions, whose precise location
is the result of an  interplay of the renormalization of
the real and the imaginary parts of the self-energy.
The emergence of these regions is a consequence of the proximity of the
density-driven Mott transition.
In the hole-doped case, the quasiparticles survive in the diagonal of the
Brillouin zone, near $(\pi/2,\pi/2)$. This state, which has a fermionic
spectrum with pointzeroes, is reminiscent
of the flux phases and can be connected smoothly to the
quasiparticles of the superconducting state.
The electron-doped case is completely different.
On a technical level, it is harder to approach the MIT
closely, and a first-order phase transition may pre-empt a continuous
approach to the insulating state.
Furthermore, the FS is renormalized towards nesting, and
the quasiparticles survive in a small region around
$(\pi,0)$ and $(0,\pi)$. These quasiparticles cannot be easily deformed into the
superconducting state as opposed to the quasiparticles which live around
$(\pi/2,\pi/2)$.
These properties of the underlying normal state of the Hubbard
model have striking resemblance to what is observed in the cuprates.
The hole-doped materials have a superconducting region  which
appears almost immediately after doping the Mott insulator. This
superconducting state  evolves continuously into the pseudogap
state which in turn evolves continuously from the Mott insulator.
In the electron-doped case, the pseudogap region, if it exists, is small,
and a much larger doping is needed to reach the superconducting
phase, which does not seem connected with the Mott insulator.
\begin{acknowledgments}
After the completion of this work, we became aware of a related work \cite{kyung}
on the $t^{\prime}=0$ case for the two dimensional Hubbard Model.
We thank K.M. Shen and Z.-X. Shen for sharing their
experimental results, A-M. Tremblay, D. S\'en\'echal, T. Stanescu,
C. Castellani and A. Georges for useful discussions.
Massimo Capone acknowledges
Italian MIUR Cofin 2003 and the NSF
support under grant DMR-0096462.
\end{acknowledgments}

\newcommand{{{\PRB}}}{{{Phys. Rev. B}}}\newcommand{{{\PRL}}}{{{Phys. Rev. Lett}}}\newcommand{{{\NPB}}}{{{Nucl. Phys.}}}\newcommand{{{\RMP}}}{{{Rev. Mod. Phys.}}}\newcommand{{{\ADV}}}{{{Adv. Phys.}}}


\begin{thebibliography}{79}
\expandafter\ifx\csname natexlab\endcsname\relax\def\natexlab#1{#1}\fi
\expandafter\ifx\csname bibnamefont\endcsname\relax
\def\bibnamefont#1{#1}\fi
\expandafter\ifx\csname bibfnamefont\endcsname\relax
\def\bibfnamefont#1{#1}\fi
\expandafter\ifx\csname citenamefont\endcsname\relax
\def\citenamefont#1{#1}\fi
\expandafter\ifx\csname url\endcsname\relax
\def\url#1{\texttt{#1}}\fi
\expandafter\ifx\csname urlprefix\endcsname\relax\def\urlprefix{URL }\fi
\providecommand{\bibinfo}[2]{#2}
\providecommand{\eprint}[2][]{\url{#2}} 
\bibitem{anderson}\bibinfo{author}{\bibfnamefont{P.~W. Anderson}}, \bibinfo{journal}{Science} \textbf{\bibinfo{volume}{235} }\bibinfo{page}{1196} (\bibinfo{year}{1987}).
\bibitem{gross-arun} \bibinfo{author}{\bibfnamefont{C. Gros}}, \bibinfo{author}{\bibfnamefont{R. Joint}} and  \bibinfo{author}{\bibfnamefont{T.~M. Rice}}, \bibinfo{journal}{ \PRB  } \textbf{\bibinfo{volume}{ 36  } }\bibinfo{page}{381} (\bibinfo{year}{1997}); \bibinfo{author}{\bibfnamefont{A. Paramekanti}}, \bibinfo{author}{\bibfnamefont{M. Randeria}} and  \bibinfo{author}{\bibfnamefont{N. Trivedi}}, \bibinfo{journal}{ \PRL  } \textbf{\bibinfo{volume}{ 87  } }\bibinfo{page}{217002} (\bibinfo{year}{2001}).
\bibitem{bza-liu} \bibinfo{author}{\bibfnamefont{G. Baskaran}}, \bibinfo{author}{\bibfnamefont{Z. Zou}} and  \bibinfo{author}{\bibfnamefont{P.~W. Anderson}}, \bibinfo{journal}{ Solid State Com.  } \textbf{\bibinfo{volume}{63 } }\bibinfo{page}{973} (\bibinfo{year}{1987}); \bibinfo{author}{\bibfnamefont{G. Kotliar}} and  \bibinfo{author}{\bibfnamefont{J. Liu}}, \bibinfo{journal}{ \PRB   } \textbf{\bibinfo{volume}{ 38  } }\bibinfo{page}{5142} (\bibinfo{year}{1988}).
\bibitem{cdmft}\bibinfo{author}{\bibfnamefont{G. Kotliar}}, \bibinfo{author}{\bibfnamefont{S.~Y. Savrasov}}, \bibinfo{author}{\bibfnamefont{G. Palsson}} and  \bibinfo{author}{\bibfnamefont{G. Biroli}}, \bibinfo{journal}{\PRL } \textbf{\bibinfo{volume}{87 } }\bibinfo{page}{186401} (\bibinfo{year}{2001}).
\bibitem{jarrellsc-lich} \bibinfo{author}{\bibfnamefont{Th. Maier}}, \bibinfo{author}{\bibfnamefont{M. Jarrell}}, \bibinfo{author}{\bibfnamefont{Th. Pruschke}} and  \bibinfo{author}{\bibfnamefont{J. Keller}}, \bibinfo{journal}{ \PRL  } \textbf{\bibinfo{volume}{ 85  } }\bibinfo{page}{1524} (\bibinfo{year}{2000}); \bibinfo{author}{\bibfnamefont{A. I. Lichtenstein}} and  \bibinfo{author}{\bibfnamefont{M. I. Katsnelson}}, \bibinfo{journal}{ \PRB  } \textbf{\bibinfo{volume}{ 62  } }\bibinfo{page}{R9283} (\bibinfo{year}{2000}).
\bibitem{damasce-campu} \bibinfo{author}{\bibfnamefont{A. Damascelli}}, \bibinfo{author}{\bibfnamefont{Z.~X. Shen}} and  \bibinfo{author}{\bibfnamefont{Z. Hussain}}, \bibinfo{journal}{\RMP} \textbf{\bibinfo{volume}{75 } }\bibinfo{page}{473} (\bibinfo{year}{2003}); "Physics of Superconductors II" \bibinfo{author}{\bibfnamefont{J. C. Campuzano}}, \bibinfo{author}{\bibfnamefont{M. R Norman}} and  \bibinfo{author}{\bibfnamefont{M. Randeria}}, {\sl K. H. Bennemann and J. B. Ketterson } (\bibinfo{year}{2004}) \bibinfo{note}{167-273}.
\bibitem{bpk}\bibinfo{author}{\bibfnamefont{O. Parcollet}}, \bibinfo{author}{\bibfnamefont{G. Biroli}} and  \bibinfo{author}{\bibfnamefont{G. Kotliar}}, \bibinfo{journal}{\PRL } \textbf{\bibinfo{volume}{92 } }\bibinfo{page}{226402} (\bibinfo{year}{2004}).
\bibitem{jarrell}\bibinfo{author}{\bibfnamefont{Th. A. Maier}}, \bibinfo{author}{\bibfnamefont{Th. Pruschke}} and  \bibinfo{author}{\bibfnamefont{M. Jarrell}}, \bibinfo{journal}{ \PRB} \textbf{\bibinfo{volume}{66 } }\bibinfo{page}{075102} (\bibinfo{year}{2002}).
\bibitem{kusko}\bibinfo{author}{\bibfnamefont{C. Kusko}}, \bibinfo{author}{\bibfnamefont{R.~S. Markiewicz}}, \bibinfo{author}{\bibfnamefont{M. Lindroos}} and  \bibinfo{author}{\bibfnamefont{A. Bansil}}, \bibinfo{journal}{\PRB } \textbf{\bibinfo{volume}{66} }\bibinfo{page}{140513} (\bibinfo{year}{2002}).
\bibitem{tremblay2}\bibinfo{author}{\bibfnamefont{B. Kyung}}, \bibinfo{author}{\bibfnamefont{V. Hankevych}}, \bibinfo{author}{\bibfnamefont{A.-M. Dare}} and  \bibinfo{author}{\bibfnamefont{A.-M.S. Tremblay}}, \bibinfo{journal}{\PRL} \textbf{\bibinfo{volume}{93 } }\bibinfo{page}{147004} (\bibinfo{year}{2004}).
\bibitem{tremblay}\bibinfo{author}{\bibfnamefont{D. S\'en\'echal}} and  \bibinfo{author}{\bibfnamefont{A-M.~S. Tremblay}}, \bibinfo{journal}{\PRL } \textbf{\bibinfo{volume}{92 } }\bibinfo{page}{126401} (\bibinfo{year}{2004}).
\bibitem{tohyama4}\bibinfo{author}{\bibfnamefont{T. Tohyama}} and  \bibinfo{author}{\bibfnamefont{S. Maekawa}}, \bibinfo{journal}{\PRB } \textbf{\bibinfo{volume}{49 } }\bibinfo{page}{3596} (\bibinfo{year}{1994}).
\bibitem{revmodmft}\bibinfo{author}{\bibfnamefont{A. Georges}}, \bibinfo{author}{\bibfnamefont{G. Kotliar}}, \bibinfo{author}{\bibfnamefont{W. Krauth}} and  \bibinfo{author}{\bibfnamefont{M. J. Rozenberg}}, \bibinfo{journal}{ \RMP} \textbf{\bibinfo{volume}{68 } }\bibinfo{page}{13} (\bibinfo{year}{1996}).
\bibitem{marce}\bibinfo{author}{\bibfnamefont{M. Capone}}, \bibinfo{author}{\bibfnamefont{M. Civelli}}, \bibinfo{author}{\bibfnamefont{S.~S.~Kancharla}}, \bibinfo{author}{\bibfnamefont{C. Castellani}} and  \bibinfo{author}{\bibfnamefont{G. Kotliar}}, \bibinfo{journal}{\PRB } \textbf{\bibinfo{volume}{69 } }\bibinfo{page}{195105} (\bibinfo{year}{2004}).
\bibitem{krauth-caffarel}\bibinfo{author}{\bibfnamefont{M. Caffarel}} and  \bibinfo{author}{\bibfnamefont{W. Krauth}}, \bibinfo{journal}{\PRL} \textbf{\bibinfo{volume}{72 } }\bibinfo{page}{1545} (\bibinfo{year}{1994}).
\bibitem{Shen}\bibinfo{author}{\bibfnamefont{K. M. Shen et al.}}, \eprint{preprint and private communication.}.
\bibitem{affgk1} \bibinfo{author}{\bibfnamefont{I. Affleck}} and  \bibinfo{author}{\bibfnamefont{B. Marston}}, \bibinfo{journal}{\PRB } \textbf{\bibinfo{volume}{37 } }\bibinfo{page}{3774} (\bibinfo{year}{1998}); \bibinfo{author}{\bibfnamefont{G. Kotliar}}, \bibinfo{journal}{\PRB } \textbf{\bibinfo{volume}{37 } }\bibinfo{page}{3664} (\bibinfo{year}{1998}).
\bibitem{kyung}\bibinfo{author}{\bibfnamefont{B. Kyung et al.}}, \bibinfo{note}{private communication}.
 \end{thebibliography}
\end{document}